\documentclass[
prb,
twocolumn,
showpacs,
groupedaddress,
superscriptaddress,
]{revtex4-1}

\usepackage{soul}
\usepackage{amsfonts,amstext}
\usepackage{graphicx}
\usepackage{dcolumn}
\usepackage{bm}
\usepackage[flushleft]{threeparttable}
\usepackage{url}
\usepackage{amsmath}
\usepackage{amssymb}
\usepackage{braket}
\usepackage{multirow}
\usepackage{array}
\usepackage{url}
\usepackage{booktabs}
\usepackage{float}
\usepackage{lineno,hyperref}
\usepackage{mathtools}
\usepackage{xcolor}

\hypersetup{colorlinks=true, urlcolor=blue, citecolor=cyan, pdfborder={0 0 0}}

\newcommand{\veck}{\boldsymbol{k}}
\newcommand{\vecr}{\boldsymbol{r}}

\newcolumntype{P}[1]{>{\centering\arraybackslash}p{#1}}
\newcommand{\red}[1]{{\color{black}{{#1}}}} 
\newcommand{\nn}{\nonumber}
\newcommand{\fig}[1]{Fig.\,\ref{#1}}
\newcommand{\figs}[1]{Figs.\,\ref{#1}}
\newcommand{\figr}[1]{Figure\,\ref{#1}}
\newcommand{\figrs}[1]{Figures\,\ref{#1}}

\usepackage{MnSymbol} 
\usepackage{enumitem,lipsum}

\usepackage{setspace}

\begin{document}

\setstcolor{blue} 

\title{\red{Ferromagnetic nodal-line metal in monolayer {\em h}-InC}}
\author{Sunam Jeon}
\affiliation{Department of Energy Science, Sungkyunkwan University (SKKU), Suwon 16419, Korea}
\author{Yun-Tak Oh}
\affiliation{Department of Physics, Sungkyunkwan University (SKKU), Suwon 16419, Korea}
\author{Youngkuk Kim}
\affiliation{Department of Physics, Sungkyunkwan University (SKKU), Suwon 16419, Korea}
\date{\today}

\begin{abstract} 
Based on first-principles calculations, we predict a new two-dimensional ferromagnetic material that exhibits exotic Fermi surface topology.  We show that monolayer hexagonal indium carbide ({\em h}-InC) is thermodynamically and dynamically stable, and it energetically favors the ferromagnetic ordering of spins.  The perfectly planar geometry in two dimensions, together with ferromagnetism, gives rise to a unique opportunity to encounter intriguing electronic properties, captured in the Fermi surface and band topology.  We show that multiple nodal lines coexist in momentum space, accompanied by the electron and hole pockets that touch each other linearly at the nodal lines.   Inclusion of spin-orbit coupling enriches the magnetic and electronic properties of {\em h}-InC.  Spin-orbit coupling leads to an easy-plane type magnetocrystalline anisotropy, and the nodal lines can be tuned into topological nodal points, contingent upon the magnetization direction.  Symmetry analysis and a tight-binding model are provided to explain the nodal structure of the bands.  Our findings suggest {\em h}-InC as a new venue for supporting carbon-based magnetism and exotic band topology in two dimensions.
\end{abstract}
\pacs{}
\maketitle

\section{INTRODUCTION}

\red{
A striking consequence of symmetry and topology  in the electronic structure of materials is the occurrence of a symmetry-protected topological state, such as topological insulators\,\cite{RevModPhys.82.3045, RevModPhys.83.1057}. Notably, the protected degenercies of electronic energy bands by nontrivial band topology  are of particular interest, leading to the discovery of condensed-matter realizations of fundamental particles, such as Majorana fermions \cite{kitaev2001unpaired, fu2008superconducting, lutchyn2010majorana}, Weyl fermions \cite{wan2011topological, burkov2011weyl}, and Dirac fermions \cite{young2012dirac, wang2012dirac}. Topological materials can also realize more exotic fermions beyond 
the conventional form of elementary particles \cite{soluyanov2015type, wieder2016double, Bradlynaaf5037}, such as nodal line semimetals \cite{burkov2011topological, Kim15p036806, yu2015topological}. Nodal line semimetals are a class of topological semimetals that are characterized by hosting one-dimensional lines of nodes formed from the conduction and valance bands. There are multiple mechanisms that topologically protect nodal lines associated with different symmetries, including inversion \cite{Kim15p036806, ahn2018band}, mirror \cite{Bian2016}, and nonsymmorphic glide-mirror~\cite{carter2012semimetal, fang2015topological} symmetries. More recently, topological semimetals without time-reversal symmetry hosted in magnetic materials have spurred a deal of community interest in finding exotic phenomena, such as tunable nodal points \cite{chang2018magnetic}, enhanced anomalous quantum Hall effect \cite{burkov2014anomalous, kim2018large}. }

\red{On the other hand, }the recent development of synthesis and characterization techniques for two-dimensional (2D) materials\,\cite{novoselov2005two, zhang2005experimental, C7CS00210F, wang2012electronics, chhowalla2013chemistry, jariwala2014emerging, yun2012thickness, zhu2011giant} has led to the exciting discovery of a room-temperature ferromagnet in two dimensions\,\cite{griffiths1964peierls, mermin1966absence, de2001experiments, vaz2008magnetism}. Whereas magnetism is one of the oldest phenomena observed in materials, its realization in a 2D system has remained elusive; it is only recently that convincing  evidence of ferromagnetism has been observed in 2D materials, which include thin films of chromium germanium telluride Cr$_2$Ge$_2$Te$_6$\,\cite{Gong2017} and chromium triiodide CrI$_3$\,\cite{Huang2017}.  The discovery of these 2D magnetic materials has motivated numerous experimental and theoretical studies\,\cite{liu2017critical, miao20182d, bonilla2018strong, lado2017origin, lin2017tricritical, kuklin2018two}, which have potential implications for future spintronic device applications\,\cite{Wolf2001, sander20172017, Yingjie17p6b04884, Hirohata15p7160747} and fundamental science\,\cite{RevModPhys.90.015001, Zhao16p195104, liu2008quantum, yu2010quantized, chang2013experimental, Fert2013, nayak2016large, Tang16p1100, watanabe2017structure, sun2018prediction, hua2018dirac, zhang2018strong}. 

Apart from these efforts, there have been ongoing studies regarding metal-free magnetism using the localize $p_z$ orbitals of carbon allotropes. According to  Lieb's theorem\,\cite{PhysRevLett.62.1201}, the ground electronic state of a bipartite lattice is accompanied by the electron spin of $s = |N_{\mathcal A} - N_{\mathcal B}|/2$, where $N_{\mathcal A}$ and $N_{\mathcal B}$ are the number of sites in the $\mathcal{A}$ and $\mathcal{B}$ sublattices, respectively. The crucial requirement for a carbon-based bipartite lattice to \red{induce  ferromagnetism} is thus to unbalance the number of the $p_z$ orbitals between the two sublattices. This condition enforces two electrons to occupy a single $\pi$-state near the Fermi level, paying an extra Coulomb repulsion energy $U$, and necessitating spin polarization depending on the strength of $U$ with respect to the hopping strength $t$ of the electrons\,\cite{PhysRevLett.62.1201, PhysRev.115.2, ANDERSON1196}.  Considered as possible realizations of Lieb's condition, diverse carbon allotropes have been investigated based on a variety of schemes\,\cite{zhou2009ferromagnetism, nair2012spin, tuvcek2016sulfur, blonski2017doping, dai2014electronic, vcervenka2009room, enoki2007electronic, son2006energy, wu2009magnetic, wakabayashi1998spin, el2003structure, ma2004magnetic, krasheninnikov2006bending, singh2009magnetism, dai2011first, Localized_Magnetic_G, tuvcek2017room, tuvcek2018emerging, son2006half, jung2009theory, nair2012spin, kopelevich2000ferromagnetic, esquinazi2002ferromagnetism, esquinazi2003induced, liu1998fullerene, saito1998physical,gonzalez2016atomic}. 

\begin{figure}
    \centering
    \includegraphics[width=0.47\textwidth]{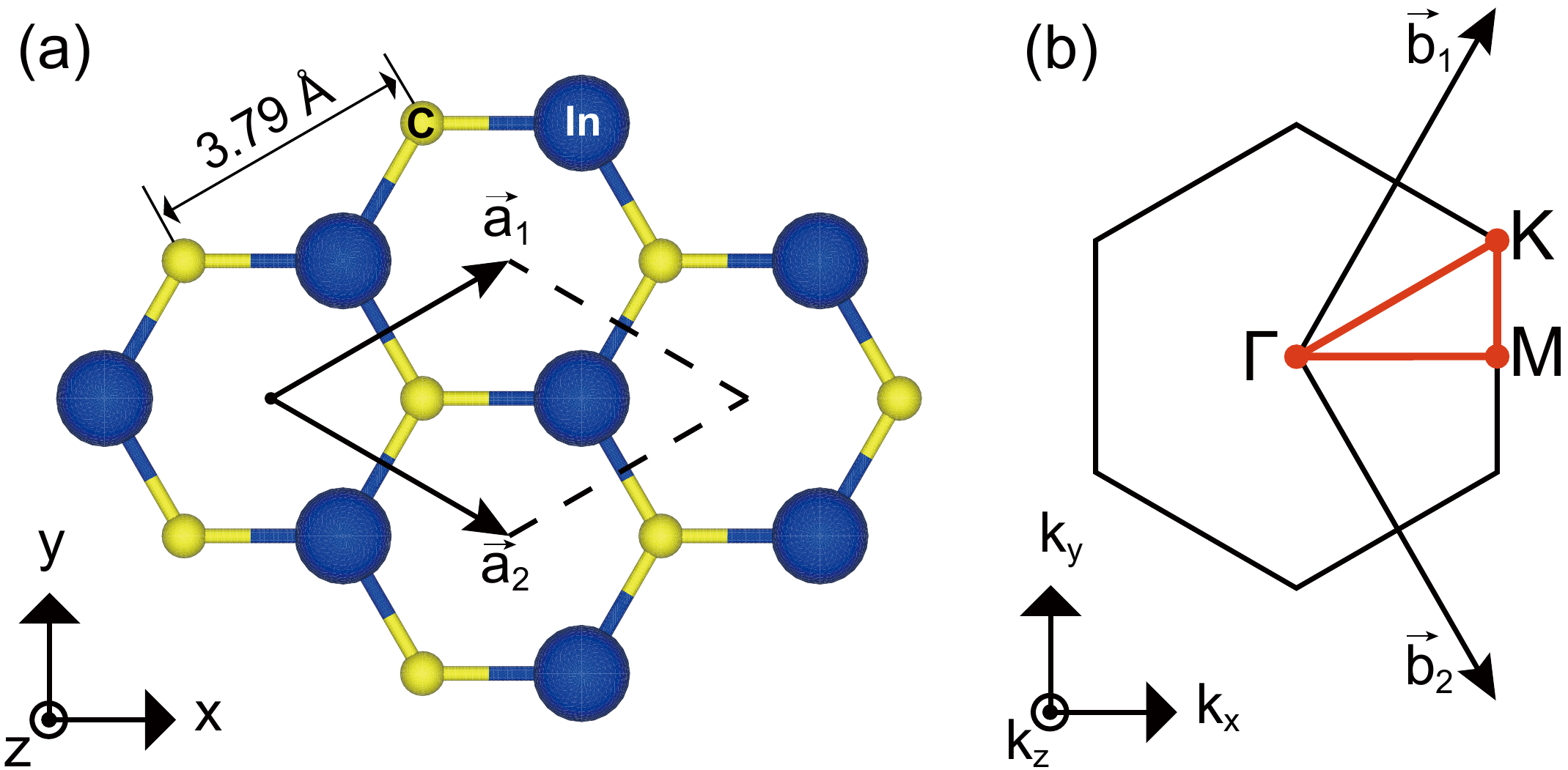}
    \caption{(a) Atomic structure of \emph{h}-InC in a planar 2D honeycomb lattice. The primitive unit cell is indicated by black dashed lines. (b) High-symmetry $\veck$ points in the 2D hexagonal Brillouin zone (BZ). }
    \label{fig:atomic}
\end{figure}

\red{
In this paper we employ the design principle for  the bipartite magnetism dictated from Lieb's theorem to find a new ferromagnetic nodal-line material in two dimensions. We suggest that a single layer of hexagonal indium carbide (\emph{h}-InC), shown in \fig{fig:atomic}, should fulfill the Lieb's condition. Using first-principles calculations, we show that {\em h}-InC is a stable ferromagnetic metal, which hosts symmetry-protected twofold-degenerate nodal lines in momentum space, referred to as Weyl nodal lines.  Based on the ferromagnetic phase of {\em h}-InC, we provide first-principles calculations that  establish {\em h}-InC  as a promising platform for realizing exotic  fermiology and potential band topology. In the absence of spin-orbit coupling (SOC), the perfectly planar structure of {\em h}-InC allows for the existence of Weyl nodal lines protected by mirror symmetry. We find that different types of nodal lines, respectively referred to as type-I and type-II nodal lines\,\cite{soluyanov2015type, type2_nodal_loop_2017, He17p1709.08287, kim1807.08523, feng2017experimental, He17p1709.08287}, coexist near the Fermi level. The Weyl nodal lines are accompanied with the electron and hole pockets that touch each other linearly at the nodal line. SOC can gap out the Weyl nodal line, or tune to a pair of nodal points, depending on magnetization orientation. The nodal points are topological, characterized by the $\pi$ Berry phase. Considered as a 2D  descendant of a three-dimensional Weyl node, which is described by a massless two-band Weyl-like Hamiltonian, the nodal point is often referred to as a 2D Weyl node. We provide symmetry analysis that offers the rationales for the presence and protection of the nodal structure, which is further supported by our a tight-binding theory. The new insight from this work could lead to the discovery of a new 2D material that allows for tunable band topology via the magnetization.}

\section{METHODS}
\label{sec:methods}
In the present study, we performed first-principles calculations based on density functional theory (DFT) as implemented in the Vienna {\it ab initio} simulation package (\textsc{VASP})\,\cite{VASP} \red{within the projector-augmented-plane-wave (PAW) method\,\cite{PAW, PAW2}}. The exchange-correlation energy was calculated within the Perdew-Burke-Ernzerholf-type generalized gradient approximation\,\cite{PBE}. The $48\times48$ of $\veck$-grid were sampled from the first Brillouin zone (BZ) using the gamma-centered Monkhorst-Pack scheme\,\cite{Monk}. The plane wave basis was constructed within the energy cutoff of 550\ eV. The two-dimensional (2D) crystal structure was approximated by placing a vacuum space of 20\ \AA\ along the out-of-plane $z$-direction. The convergence thresholds for the energy and the Hellmann-Feynman force were set to $10^{-6}$\ eV and $0.01$\ eV/\AA, respectively.  The phonon calculations were carried out by using the {\textsc{Phonopy}} package\,\cite{PHONO}. \red{The spin-polarized (spin-unpolarized) phononic energy spectra were calculated within the force criterion of $10^{-5}$\ eV/\AA\ using a 4$\times$4$\times$1 (6$\times$6$\times$1) supercell and  the 6$\times$6$\times$1 Monkhorst-Pack sampling of $\veck$-points. The cohesive energy of {\em h}-InC was calculated using $E_{\rm coh}=(E_{\rm InC}-E_{\rm C}-E_{\rm In})/N_{\rm atoms}$, where $E_{\rm InC}$, $E_{\rm C}$ and $E_{\rm In}$ are the total energies of {\em h}-InC,  C, and In, respectively. $N_{\rm atoms} = 2$ is the number of atoms per unit cell.  The magnetic anisotropy energy $E_{\rm MAE}$ of {\em h}-InC was calculated based on $E_{\rm MAE}(\theta,\phi) = \tilde{E}(\theta,\phi) - \tilde{E}(0,\pi/2)$, where $\tilde{E}(\theta,\phi)$ is the total energy evaluated by constraining the direction of the magnetic moments along $\hat{\boldsymbol{n}}=(\sin\theta\cos\phi,\sin\theta\sin\phi,\cos\theta)$ within the magnetic force theorem method \cite{Liechtenstein87p65, Daalderop90p11919} implemented in \textsc{VASP}  \cite{Steiner16p224425}.   To topologically characterize the 2D Weyl points, the Berry phases were calculated via the wannierization of DFT bands using the {\sc Wannier90}\,\cite{Wannier} code.
} 

\section{Results and Discussion}
\label{sec:results}
\subsection{Atomic structure and structural stability  of {\em h}-InC}
\begin{figure}
\centering
\includegraphics[width=0.48\textwidth]{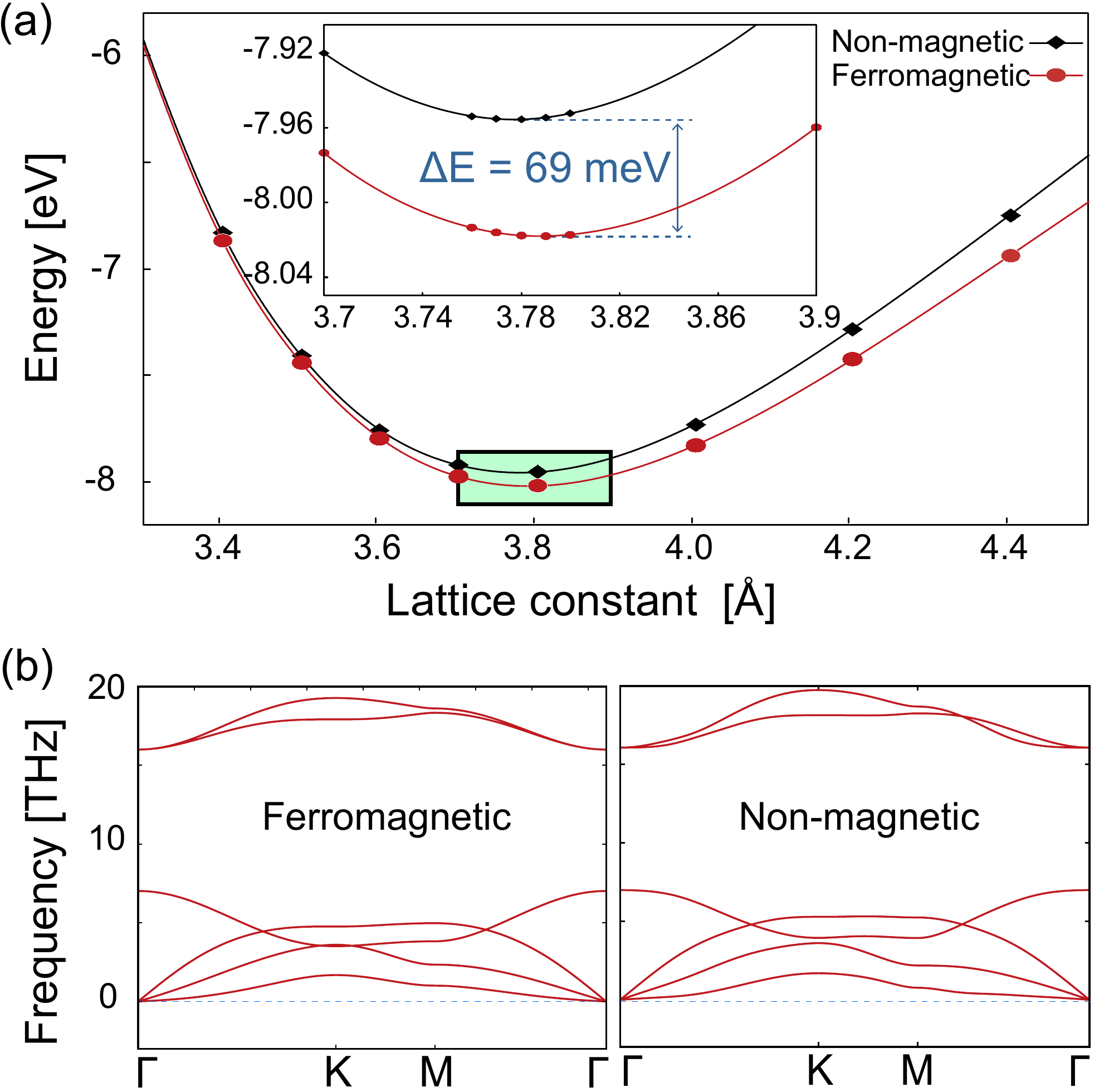}
    \caption{(a) Total energy of \emph{h}-InC evaluated as a function of the unit cell area. (Inset) Magnified view of the energy curves near the energy minima, highlighted by a green box. (b) Phononic energy spectra calculated under the ferromagnetic (left) and non-magnetic (right) spin configurations.  }
\label{fig:stability}
\end{figure}

We begin our study by delineating the crystal structure of \emph{h}-InC. Geometry optimization leads to the planar 2D honeycomb structure shown in \fig{fig:atomic}(a) with the primitive unit cell comprising one InC formula unit. The space group of \emph{h}-InC is $P\overline{6}m2$ (\# 187). The generating point group of $P\overline{6}m2$ is isomorphic to $D_{3h}^1$, which contains mirror symmetry $M_z$ \red{($M_y$)} about the $x$-$y$ \red{($x$-$z$)} plane. As shown in detail later, $M_z$ \red{($M_y$)} plays an important role to protect nodal lines \red{(nodal points)} of the electronic energy bands. The lattice constant is calculated as $a = |\mathbf{a}_1| = |\mathbf{a}_2|$ = 3.79\ {\AA}. Note that the nearest-neighboring C atoms  are separated by 3.79\ {\AA}, \red{which is much greater than the case of graphene (1.42\ \AA). This large spacing between carbons helps electrons localized at the C $p_z$ orbitals, thereby promoting spin-polarization. }

\begin{figure*}
    \centering
    \includegraphics[width=0.83\textwidth]{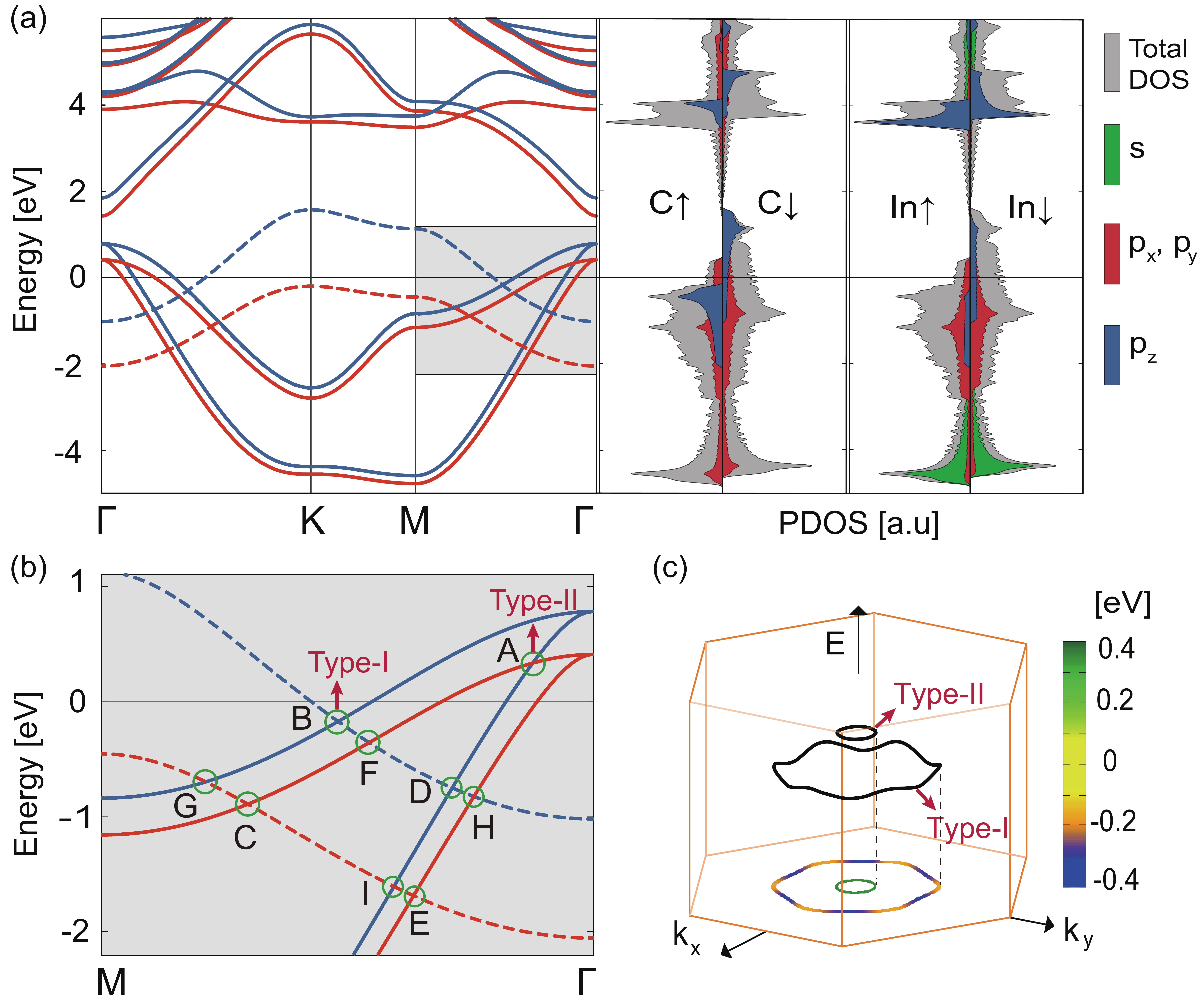}
    \caption{\red{(a) Electronic energy band structure and projected density of states (PDOS) of \textit{h}-InC. (Left panel) Spin-polarized electronic energy band structure. The Fermi level is set to zero. The energy bands are calculated without SOC along the high-symmetry $k$-points of the hexagonal BZ shown in \fig{fig:atomic}(b). The spin-up (majority spin) and spin-down (minority spin) bands are colored by red and blue, respectively. The mirror eigenbands with $m_z$ = +1 and  $m_z=$ -1 are illustrated with solid and dashed lines, respectively. (Right panel) PDOS of \emph{h}-InC. The same energy scale is used between the band structure and PDOS.  (b) Magnified view of the gray box in the left panel of (a).  A band crossing is highlighted by a green circle. (c) Weyl nodal lines in energy-momentum space. Figure plots only the nodal lines within the energy range $|E| < 0.4\ $eV. The color scheme is used to represent the energy of the Weyl nodal lines. The node that is labeled by A (B) in (b) is a member of  the type-II (type-I) Weyl nodal line in (c).}}
    \label{fig:es}
\end{figure*}

\red{We expect that the {\em h}-InC lattice is thermodynamically and kinetically stable. The reason for this follows.  First, the cohesive energy  of {\em h}-InC is calculated as -3.91\ eV per atom. This value is comparable to those of known 2D materials, such as silicene (-3.96\ eV), germanene (-2.6 \ eV) \cite{Ding2015}, as well as existing monolayer transition metal dichalcogenides, including MoTe$_2$ and WSe$_2$ (-4 $\sim$ -6 \ eV) \cite{Kang2013, Ataca2011, Fan2015, Torun2016, Kan2015}. Moreover, the {\em h}-InC lattice forms a local minimum in atomic configuration space as shown in the total energy curve in \fig{fig:es}(a), which supports that {\em h}-InC is thermodynamically stable. In addition to the thermodynamic stability, the kinetic stability of {\em h}-InC is also expected from our phononic energy calculations.  \figr{fig:stability}(b) shows the phononic energy spectra calculated with and without spin-polarization. Both results produce no negative mode, irrespective of spin-polarization, which establish the structural stability of {\em h}-InC.}

\begin{figure*}
    \centering
    \includegraphics[width=0.95 \textwidth]{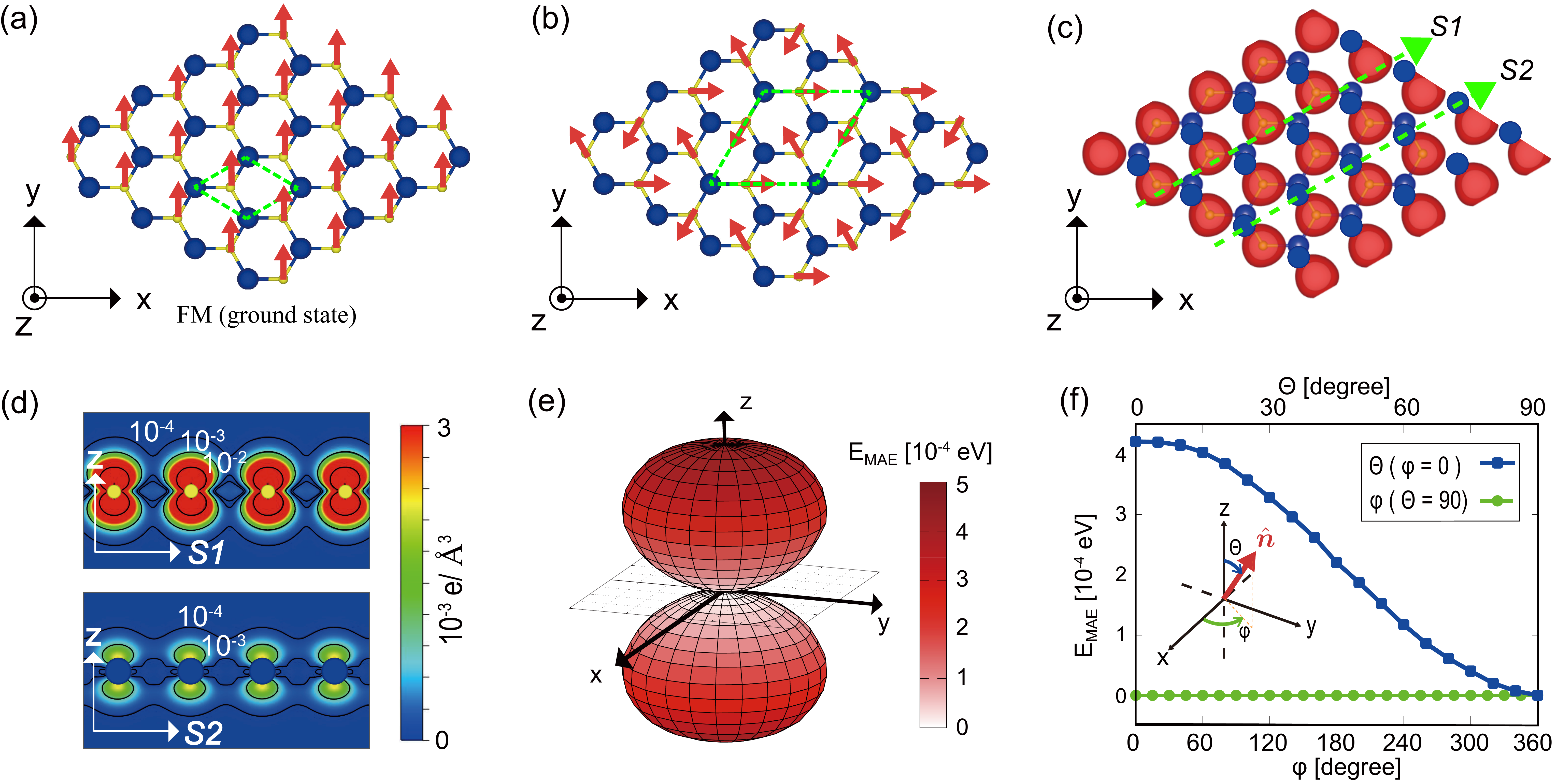}
    \caption{(a) Ferromagnetic spin configuration aligned along the in-plane $y$-direction. The corresponding unit cells are represented by green-dashed rhombuses. (b) Coplanar 120$^\circ$ spin configuration. (c)  Magnetization density isosurface at  $n_\uparrow(\vecr)-n_\downarrow(\vecr) = 0.001\ e$/\AA$^3$. (d) Cross-sectional views of the magnetization density plotted along the $S1$ (top panel) and $S2$ (bottom panel) green-dashed lines introduced in (c). \red{(e) Uniaxial magnetocrystalline anisotropic energy  $E_{\rm MAE}$ drawn in polar coordinates ($r,\theta,\phi$), where $r = E_{\rm MAE}(\theta,\phi)$. (f)  Uniaxial magnetocrystalline anisotropic energy curves $E_{\rm MAE} = E_{\rm MAE} (\theta,\phi)$ evaluated as a function of $\theta$ at $\phi = 0$  (blue) and as a function of $\phi$ at $\theta = \pi/2$ (green), respectively. Inset illustrates the parametrization of spin orientation in terms of the azimuth angle $\theta$ measured from the out-of-plane axis and the horizontal azimuth angle $\phi$ measured on the basal plane. The red arrow represents spin orientation $\hat{\boldsymbol{n}}=(\sin\theta\cos\phi,\sin\theta\sin\phi,\cos\theta)$ in real space. }
}
 \label{fig:mag}
\end{figure*}
\subsection{Magnetism of {\em h}-InC}
\red{
Having discussed the atomic structure of {\em h}-InC, we now turn to its magnetic properties. We first evidence that {\em h}-InC has the ferromagnetic ground state. After many trials  to minimize the total energy of the system with different spin configurations, including the coplanar 120$^\circ$ antiferromagnetic configuration shown in \fig{fig:mag}(b), we find that the ferromagnetic spin configuration in \fig{fig:mag}(a) results in the lowest total energy. In detail, the total energy per unit cell of the ferromagnetic state is calculated as 69\ meV lower per carbon atom than the non-magnetic state [See \fig{fig:stability}(a)]. With respect to the coplanar 120$^\circ$ antiferromagnetic spin configuration shown in the right panel of \fig{fig:mag}(a), the energy of the ferromagnetic state is 34\ meV lower per carbon atom. As we will show below, spins are localized at carbon atoms that form a triangular sublattice, in which geometric frustration forbids the colinear 180$^\circ$ antiferromagnetic spin configuration to become a ground spin state.}

\red{
We explain the ferromagnetic ground state of {\em h}-InC as being due to the unbalanced $p_z$ electrons. As an element in group XIII (main group III) of the periodic table, In provides one less valence electrons than C per unit cell.  In this respect, {\em h}-InC can be viewed as In-substituted graphene, in which one sublattice of graphene is substituted with In atoms. The In-substitution plays a role to remove half of the $p_z$ electrons of graphene, which necessitates the spin-polarization fulfilling Lieb's magnetization condition. Indeed, the net spin moment of the ferromagnetic ground state is calculated as 0.8\,$\mu_\mathrm{B}$ per unit cell, which is close to 1 $\mu_\mathrm{B}$ that we expected from Lieb's theorem.   Moreover, the magnetization density, shown in \fig{fig:mag}(c), shows that a large amount of spins are  localized at the carbon sites, which indicate that carbon atoms mainly contribute to the spin moment.  The cross-sectional views of the magnetization density in \fig{fig:mag}(d) further reveal that the magnetization density originates from the electrons in the carbon $p_z$ orbitals, featuring the $p_z$ orbital characters. Therefore, carbon-based magnetism is considered to be at the heart of the magnetism of {\em h}-InC.
}

\red{
In the presence of SOC, {\em h}-InC exhibits a weak magnetocrystalline anisotropy of an easy-plane type. \figr{fig:mag}(e) shows the magnetic anisotropy energy $E_{\rm MAE}$ of {\em h}-InC evaluated as a function of the azimuth angle $\theta$ measured from the out-of-plane axis and the horizontal azimuth angle $\phi$ measured on the basal plane [See inset of \fig{fig:mag}(f)]. The result shows that $E_{\rm MAE}$ has the minimum at $\theta = \pi/2$ with a negligible dependence on $\phi$, thus leading to the formation of (0001) easy plane and in-plane isotropic. The (out-of-plane) hard  axis energy is amount to 0.4\ meV, as shown in \fig{fig:mag}(f). The in-plane isotropy could make the ferromagnetic phase susceptible to thermal fluctuation \cite{PhysRevLett.17.1133} and correspondingly difficult to realize a 2D magnet at finite temperature using {\em h}-InC.}

\subsection{Electronic structure of {\em h}-InC}
\subsubsection{Electronic band structure and PDOS without SOC}

Having  predicted the existence of a new 2D magnetic material, we turn to the characterization of its electronic structure.  It is readily noticed that \emph{h}-InC is a spin-polarized metal from its electronic energy band structure presented in \fig{fig:es}(a). The Fermi level ($E = 0$) intersects with the spin-polarized energy bands. Notably, the largest spin-splitting occurs between the C $p_z$ bands that correspond to the energy bands containing the two highest occupied energy levels at the $\Gamma$ point, depicted by dashed lines in \figs{fig:es}(a) and \ref{fig:es}(b). The projected density of states (PDOS) in the right panel of \fig{fig:es}(a) further reveal that these energy bands mainly consist of the C $p_z$ orbitals, giving rise to the strong PDOS peaks at $E = -0.430$\ eV from the spin-up band and at $E = 1.161$\ eV from the spin-down band. As expected, these C $p_z$ bands have relatively a narrow bandwidth within $\sim$1.8\ eV, reflecting the hindered electron hopping between the C $p_z$ orbitals due to the interstitial In atoms.

The PDOS further reveal that all the other energy bands except the C $p_z$ bands near the Fermi level  consist of the $s$, $p_x$, and $p_y$ orbitals from both C and In atoms. In detail, the bonding states of the $s$ orbitals are located deep inside the occupied region of the energy bands near $E = $ -10\ eV, whereas the anti-bonding states of the $s$-orbitals are placed above the Fermi level near $E =$ 1.43\ eV and 1.9\ eV at the $\Gamma$ point. In addition, the strong PDOS peaks appear right below the Fermi level, contributed from the $p_x$ and $p_y$ orbitals of both C and In as well as with a sizable contribution from the In and C $s$ orbitals. This indicates that $\sigma$ bonds are formed between the $sp2$-type orbitals of In and C. Hereafter we refer to these bands as $sp2$ bands. Note that the $sp2$ ($p_z$) bands are symmetric (anti-symmetric) under the mirror operation $M_z$, as they are formed from the $s$, $p_x$, and $p_y$ ($p_z$) orbitals.

\begin{table}[h!]
\red{
    \centering
    \begin{tabular}{|c|c|c|c|c|c|c|}
    \hline
     Node  & ($m_z,s_z$) &  ($m_z,s_z$) & nodal line\\
    \hline
     A     &  ($+$ , $-$)   &  ($+$ , $+$)    & Type-II \\
     B     &  ($+$ , $-$)   &  ($-$ , $-$)      & Type-I  \\
     C     &  ($+$ , $+$)   &  ($-$ , $+$)     & Type-I  \\
     D     &  ($+$ , $-$)   &  ($-$ , $-$)      & Type-I  \\
     E     &  ($+$ , $+$)   &  ($-$ , $+$)     & Type-I  \\
     F     &  ($+$ , $+$)   &  ($-$ , $-$)      & Type-I  \\
     G    &  ($+$ , $-$)    &  ($-$ , $+$)     & Type-I \\
     H    &  ($+$ , $+$)   &  ($-$ , $-$)      & Type-I \\
     I     &  ($+$ , $-$)    &  ($-$ , $+$)      & Type-I \\  
    \hline
 \end{tabular}
  \caption{Mirror ($m_z=\pm 1$) and spin ($s_z=\pm 1$) eigenvalues for the energy bands without SOC that form the corresponding nodes defined in \fig{fig:es}(b). The second (third) column lists the symmetry eigenvalues for the bands that are located on the left upper (lower) side of the node in \fig{fig:es}(b). The type-I and type-II classification of the Weyl nodal lines is listed in the fourth column.}
    \label{table:stable1}
    }
\end{table}

The energy bands with different mirror eigenvalues ($m_z={\pm}1$) can cross each other without opening a band gap as shown in \figs{fig:es}(a) and \ref{fig:es}(b). A mirror $+$ band ($m_z =  +1$, solid line) intersects with a mirror $-$ band ($m_z = -1$, dashed line). Similarly, the energy bands with different spin states ($s_z  = {\pm}1$)  can also overlap each other without opening a band gap; a spin-up band ($s_z=+1$, colored by red) crosses a spin-down band ($s_z=-1$, colored by blue). Thereby, the mirror and  spin-rotational $SU(2)$ symmetries independently protect the nodal structure of the energy bands, and all the band crossings in \fig{fig:es}(c) are explained based on the mirror and spin eigenvalues. For example,  protected by mirror symmetry $M_z$, the band crossings at B, C, D, and E in \fig{fig:es}(c) are formed from the energy bands with the same spin state.  \red{Table\ \ref{table:stable1} lists the mirror and spin eigenvalues for the bands that form the nodal points that are introduced in \fig{fig:es}(b).}

\begin{figure*}
\centering
    \includegraphics[width=0.85\textwidth]{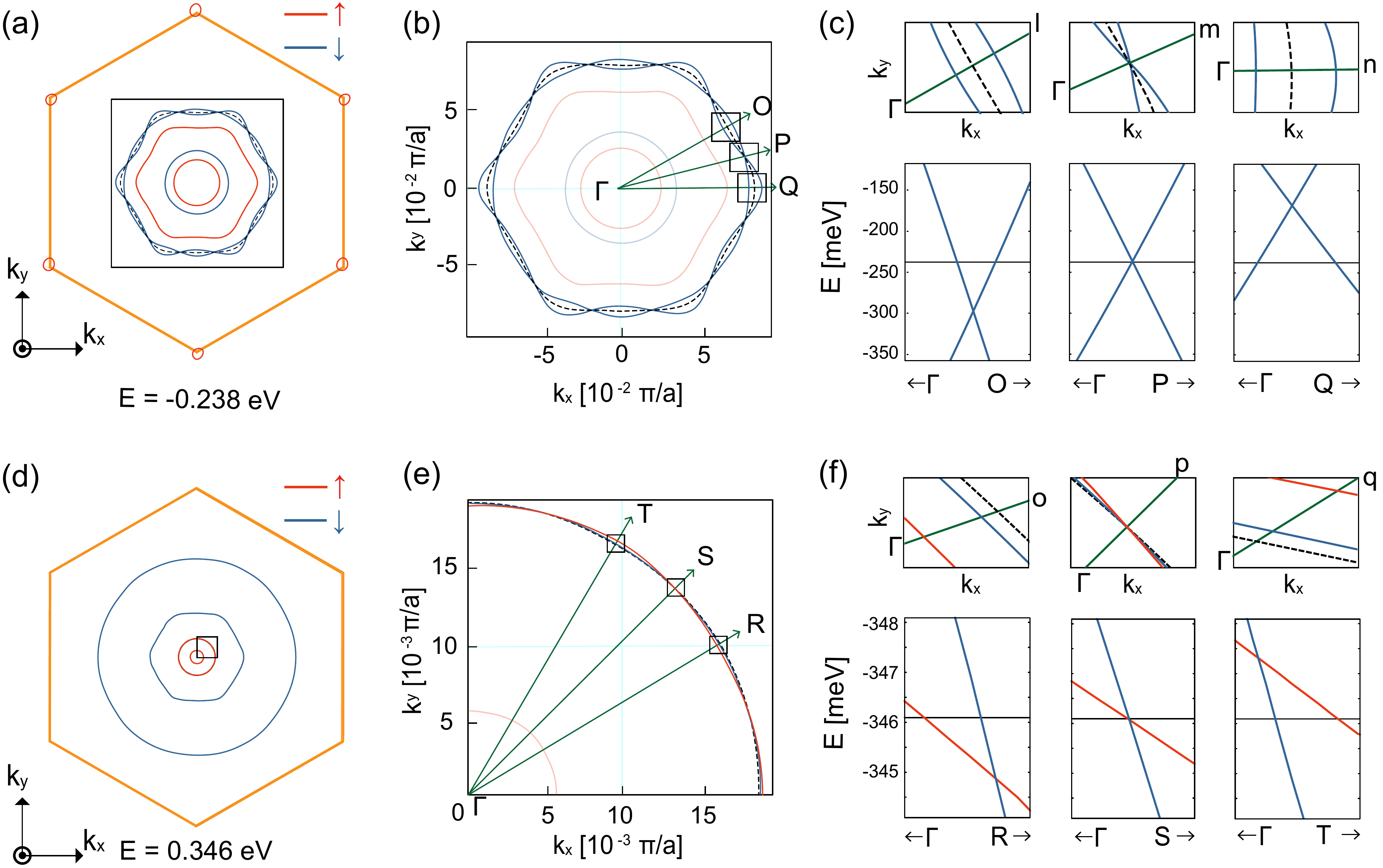}
    \caption{Contrasting Fermi surfaces of type-I and type-II nodal lines in \emph{h}-InC. (a) Fermi surfaces at $E$ = -0.238 eV, which intersect with a type-I nodal line.  The orange hexagon illustrates the BZ. The Fermi surfaces for the spin-up and spin-down bands are colored by red and blue, respectively. The momenta of the nodal line are indicated by  the dashed line.  (b) Magnified black-boxed region from (a).  Only the Fermi surfaces (contours) that intersect with the type-I nodal line are highlighted.  (c) Magnified black-boxed regions from (b) (top panels) and the energy bands depicted along the green line from the corresponding top panel (bottom panels). (d) Fermi surfaces at $E$ = 0.346\ eV, which intersect with the type-II nodal line. (e) Magnified black-boxed region from (d). Only the Fermi surfaces (contours) that intersect with the type-II nodal line are highlighted.   (f) Magnified black-boxed regions from (e) (top panels) and the energy bands depicted along the green line from the corresponding top panel (bottom panels).}
    \label{fig:fs}
\end{figure*}

\subsubsection{Weyl nodal lines and Fermi surface topology}

Being preserved independently without SOC, the mirror $M_z$ and spin-rotational $SU(2)$ symmetries are global symmetries in momentum space in the sense that they are respected at any $\veck$ point of the entire BZ.  Therefore, the occurrence of band crossings that we found from the band structure signals the existence of one-dimensional nodal lines in the 2D BZ.  A close inspection has found multiple nodal lines, such  as shown in \fig{fig:es}(d). Notably, the nodal line that contains the  band crossing A in \fig{fig:es}(c) is of a unique type, referred to as a  type-II nodal line formed from the energy bands that have the same band velocities along the nodal lines\cite{type2_nodal_loop_2017,  kim1807.08523}. Note that the $p_z$ ($sp2$) bands have a positive (negative) band curvature at $\Gamma$.  While conventional type-I nodal lines are formed dominantly from overlapping one $p_z$ band and one $sp2$ band, a type-II nodal line coexists, formed from the overlapping two $sp2$ bands in the energy range from 0.345\ eV to 0.348\ eV.

We find a contrasting behavior between the type-I and type-II nodal lines, featured in the Fermi surface geometry.  In the type-I case, an alternating chain of electron and hole pockets appears such that the position of the nodal line is placed inside the chain; in contrast, in the type-II case, a chain of electron and hole pockets occurs avoiding the position of the nodal line. Note that in both cases, the electron and hole pockets exist and touch each other linearly at the nodal line.  \figrs{fig:fs}(a) and \ref{fig:fs}(d) illustrate the Fermi surface diagrams at $E = -0.238$\ eV  and $E = 0.346$\ eV, where type-I and type-II nodal lines occur, respectively. They demonstrate that the nodal line that is represented by a dashed line in \fig{fig:fs}(b) [(e)] is located at the interior (exterior) part of the area surrounded by the Fermi surfaces that are represented by solid contours. Note that the Fermi surfaces are warped hexagonally around the $\Gamma$ point. The hexagonal warping is more prominent in the type-I nodal line in \fig{fig:fs}(b) than the type-II nodal line in \fig{fig:fs}(e), which is due to the type-II nodal line residing closer to the $\Gamma$ point. 

We propose the contrasting geometry of the Fermi surfaces as a generic feature of the type-I and type-II nodal lines in two dimensions, originated from the hexagonal warping and the different signs of the band velocities. To demonstrate this, we plot the energy bands in momentum space along the $\Gamma-A$, $\Gamma-B$, and $\Gamma-C$ lines in \fig{fig:fs}(c). We first notice that the energy disperses along the nodal line due to the hexagonal warping. The figures show that the band crossing on the $\Gamma-A$ ($\Gamma-C$) line is shifted down (up) in energy with respect to the Fermi energy ($E = -0.238$), forming an electron (hole) pocket, whereas the band crossing point occurs at the Fermi energy on $\Gamma-B$.  Thus, from $\Gamma-A$ to $\Gamma-C$ through $\Gamma-B$, the electron pocket is reduced to the crossing point, and evolves to the hole pocket. Meanwhile, the different signs of the band velocity necessitates the occurrence of the crossing points in the inner area of the two Fermi surfaces. Similarly, from $\Gamma-D$ to $\Gamma-F$ through $\Gamma-E$ for the Fermi surfaces at $E = 0.346$\ eV, where the type-II nodal line occurs [See \fig{fig:fs}(e).], an electron pocket evolves to a hole pocket via a band crossing point.  Note that the weaker warping of the energy bands results in a weaker energy dispersion along the type-II nodal line with respect to the type-I nodal line. Moreover, unlike the type-I case, in the type-II case, the same sign of the band velocities necessitate the occurrence of the type-II crossing points in the outer area of the two Fermi surfaces as illustrated in \fig{fig:fs}(f).  These contrasting patterns of the Fermi surfaces are intrinsic to the very definition of the type-I/type-II classification, done based the band velocities, thus characterizing the types of the nodal lines in two dimensions. We expect that the characteristic feature should be observable in \emph{h}-InC as in the case of topological nodal lines in three dimensions\,\cite{mikitik1999manifestation,  mikitik2004berry,  mikitik2006band,  mikitik2007phase,  hubbard2011millikelvin, mikitik2014dirac,  HfC} and type-II Weyl nodes in topological semimetals\,\cite{Yazyev2016, Deng2016, Yu2016p}.

\subsubsection{Effect of spin-orbit coupling}

\begin{table}[h!]
\red{
    \centering
    \begin{tabular}{|c|c|c|c|c|c|c|}
    \hline
     Node  & ($m_z,s_z$) & $\mu_z$ &  ($m_z,s_z$) & $\mu_z$ & SOC gap(eV) & Nodal line\\
    \hline
     A     &  ($+$ , $-$)  &  $-$   &    ($+$ , $+$) & $+$  &  $-$  & Type-II \\
     B     &  ($+$ , $-$)  &  $-$   &    ($-$ , $-$) & $+$  &  $-$  & Type-I  \\
     C     &  ($+$ , $+$)  &  $+$   &    ($-$ , $+$) & $-$  & $-$  & Type-I  \\
     D     &  ($+$ , $-$)  &  $-$   &    ($-$ , $-$) & $+$  &  $-$  & Type-I  \\
     E     &  ($+$ , $+$)  &  $+$   &    ($-$ , $+$) & $-$  &  $-$  & Type-I  \\
     F     &  ($+$ , $+$)  &  $+$   &    ($-$ , $-$) & $+$  &  0.11 &  $-$  \\
     G     &  ($+$ , $-$)  &  $-$   &    ($-$ , $+$) & $-$  &  0.06 &  $-$ \\
     H     &  ($+$ , $+$)  &  $+$   &    ($-$ , $-$) & $+$  &  0.07 &  $-$ \\
     I     &  ($+$ , $-$)  &  $-$   &    ($-$ , $+$) & $-$  &  0.05 &  $-$ \\  
    \hline
    \end{tabular}
    \caption{Mirror and spin eigenvalues. The mirror $m_z = \pm 1$ and spin $s_z = \pm 1$ eigenvalues are evaluated from the non-SOC calculations for the corresponding nodes that were found without SOC. The mirror eigenvalues $\mu_z = \pm i$ are evaluated from the SOC calculations with the fixed spin orientation along the out-of-plane $z$-direction. The third (fifth) column lists the mirror eigenvalues of the band that appears on the left top (left bottom) side of a crossing or an anti-crossing in \fig{fig:soc_es}(a).}
    \label{table:stable2}
    }
\end{table}

\begin{figure}[h!]
\red{
    \centering
    \includegraphics[width=0.49\textwidth]{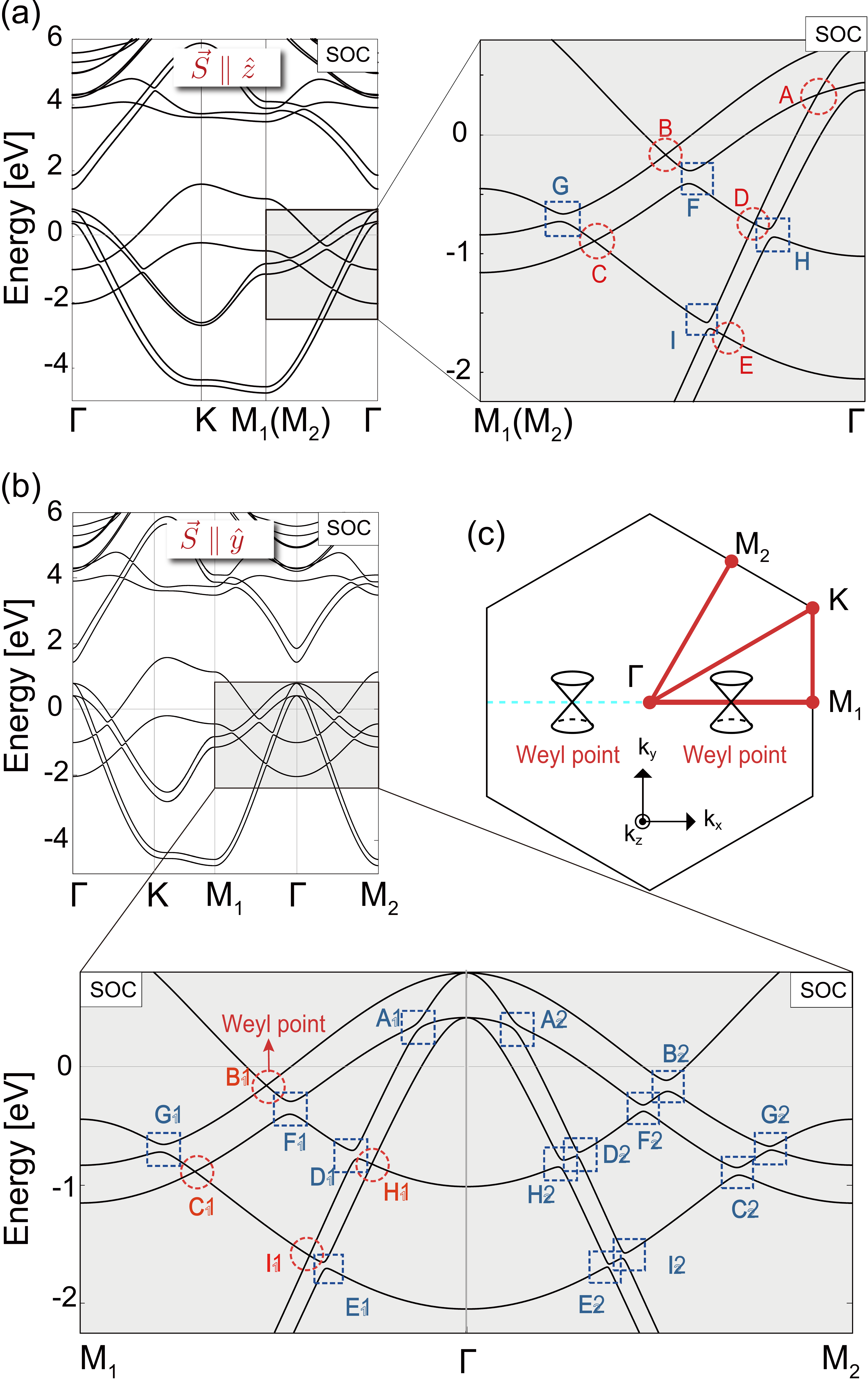}
    \caption{SOC band structures of \emph{h}-InC with different spin orientations. (a) Spins are constrained  to the out-of-plane $z$ direction. The gray-boxed region is magnified on the right panel. (b) Spins are constrained to the in-plane $y$-direction. The gray-boxed region is magnified on the bottom panel.  The red-dashed circles (blue-dashed boxes) indicate the crossing (anti-crossing) points.   (c) High-symmetry $\veck$ points of the 2D BZ. The sky-blue dashed line at $k_y = 0$ is $M_y$ invariant, where the Weyl points appear. The location of the Weyl points (labeled by B$_1$) is  marked with a Weyl cone.}
    \label{fig:soc_es}
}
\end{figure}
\red{
In general, SOC couples the spin and orbital sectors of the Bloch states, such that the nodal lines that we found without SOC can be annihilated.  The nodal lines only survive when they are protected by a global symmetry that is respected in the entire BZ even in the presence of SOC. An example is the mirror symmetry with respect to the basal plane $M_z$, which is preserved only when spins are colinearly aligned along the out-of-plane $z$-direction when including SOC. In this particular case of spin configuration, the nodal lines can survive when they are formed from the energy bands with distinct mirror eigenvalues. Inclusion of SOC necessitates the definition of a new mirror quantum number  $\mu_z = m_z  s_z = \pm i$ capable of protecting the band crossings.  Some nodal lines indeed survive in our first-principles calculations calculated by constraining the spin orientation along the out-of-plane $z$ direction as shown in \fig{fig:soc_es}(a). While the nodal lines that contain the crossings A, B, C, D, or E survive [See \fig{fig:es}(b) and \fig{fig:soc_es}(a)], the other nodal lines that contain F, G, H, or I are gaped out by including SOC. These results are in good agreement with the calculated mirror eigenvalues of the corresponding band, listed in Table\,\ref{table:stable2}.

In contrast to the out-of-plane spin-polarization, when spins are aligned uniaxially along an in-plane direction, which is the energetically most favored, the nodal lines are expected to disappear, except for the case where spins are all aligned along $y$-direction. In this particular case, the $M_y$ mirror-symmetry is preserved and the nodal lines that we found without SOC are tuned to 2D Weyl points, residing along the $M_y$ mirror symmetric $\Gamma - M_1$  line ($k_y = 0$).  The DFT band structure in \fig{fig:soc_es}(b), which is calculated by constraining spins along the $y$-direction, exhibits the Weyl points at B$_1$, C$_1$, and H$_1$ along the $M_y$-invariant $M_1 - \Gamma$ line. In contrast, all the nodes in the $\Gamma - M_2$ line, where $M_z$ is broken, open a gap. We note that the Weyl points are topological in the sense that they carry the Berry phase $\pi$. We confirmed that the Weyl point B$_1$ gives rise to the Berry phase $\pi$ when calculated along the closed path of radius 0.0038\ \AA$^{-1}$ that encircles B$_1$.}

\red{
\subsubsection{Tight-binding theory of {\em h}-InC}
Finally, we construct a tight-binding theory that describes the $s$ and $p$-orbitals of C and In of {\em h}-InC. These orbitals are responsible for the electronic structure of \emph{h}-InC near the Fermi energy. The underlying interactions that give rise to the low-energy band structure obtained from the  first-principles calculations can essentially be ascribed to 1) the kinetic energy term $\mathcal{H}_0$, contributed from both the nearest and next-nearest neighbor hopping of the valence electrons, 2) on-site energy term $\mathcal {H}_\mathrm {on-site}$ for both the electrons that are bound to each sublattice,  3) the effective Zeeman term $\mathcal{H}_\mathrm {Zeeman}$, which is constructed to discern the sites and the spins, and 4) the SOC term $\mathcal {H}_\mathrm {SOC}$, summarizing
\begin{eqnarray}
 \mathcal{H} =  \mathcal{H}_0 + \mathcal{H}_{\rm{on-site}} +  \mathcal{H}_{\rm{Zeeman}}  + \mathcal{H}_{\rm{SOC}},
\end{eqnarray}
where
\begin{eqnarray}
\mathcal{H}_{\rm{0}} \, &=& \,\sum_{j,j',l,l'} \, \sum_{\langle  \vecr,\vecr'  \rangle} \, t^{ l l'}_{1} \ {\hat{\rm{c}}^{\dagger}_{j l \alpha}(\vecr) \,{\hat{\rm{c}}_{j'  l' \alpha}(\vecr')} }  \nn \\
 &+& \, \sum_{j,l,l'} \, \sum_{\llangle \rm{\vecr,\vecr'} \rrangle} \, t_{2,j} \ \hat{\rm{c}}^{\dagger}_{j l \alpha}(\vecr) \,\hat{\rm{c}}_{j l' \alpha}(\vecr') , \nn \\
\mathcal{H}_{\rm {on-site}} \,& =& \, \sum_{\alpha}\, \sum_{j,l,\vecr} \,   \, \Delta_{j}^{l} \ \hat{\rm{c}}^{\dagger}_{j l \alpha}(\vecr) \,\hat{\rm{c}}_{j  l \alpha}( \vecr )    , \nn \\
\mathcal{H}_{\rm{Zeeman}} \,& =& - \sum_{\alpha,\beta} \, \sum_{j,l,\vecr} \, (\boldsymbol{B}^l_j \cdot \boldsymbol{s})_{\alpha \beta} \  \hat{\rm{c}}^{\dagger}_{j l \alpha}(\vecr ) \, \hat{\rm{c}}_{j  l \beta}( \vecr ),  \nn  \\
\mathcal{H}_{\rm{SOC}} \, &=& \, \sum_{\alpha,\beta,\vecr} \, i \, \lambda \,  \rm{\textbf{s}}_{\alpha \beta} \, \cdot \,\hat{\rm{\textbf{c}}}^{\dagger}_{j\alpha}( \vecr ) \, \times \,  \hat{\rm{\textbf{c}}}_{j' \beta}( \vecr )  \nn.
\end{eqnarray}
Here, $l$ and $j$ describe the orbital and sublattice indices, respectively. $\alpha$ and $\beta$ represent spin states, respectively. $t_{1}$ and $t_{2}$ are hopping parameters for the nearest and the next nearest neighbor hopping, respectively.  $\mathcal{H}_{\rm{Zeeman}}$ describes the site-dependent Zeeman splitting for $p_z$ bands and $sp2$ bands.
In the SOC Hamiltonian $\mathcal{H}_{\rm{SOC}}$, $L_{j} = i\, \varepsilon_{jkl}$ is the orbital angular momentum in the $p$-orbital basis. $\hat{\textbf{c}}_{j\alpha}(\vecr)$ describes the $p$-orbitals for spin $\alpha$ and sublattice $j$. 

\begin{figure} [h!]
\red{
    \includegraphics[width=0.5\textwidth]{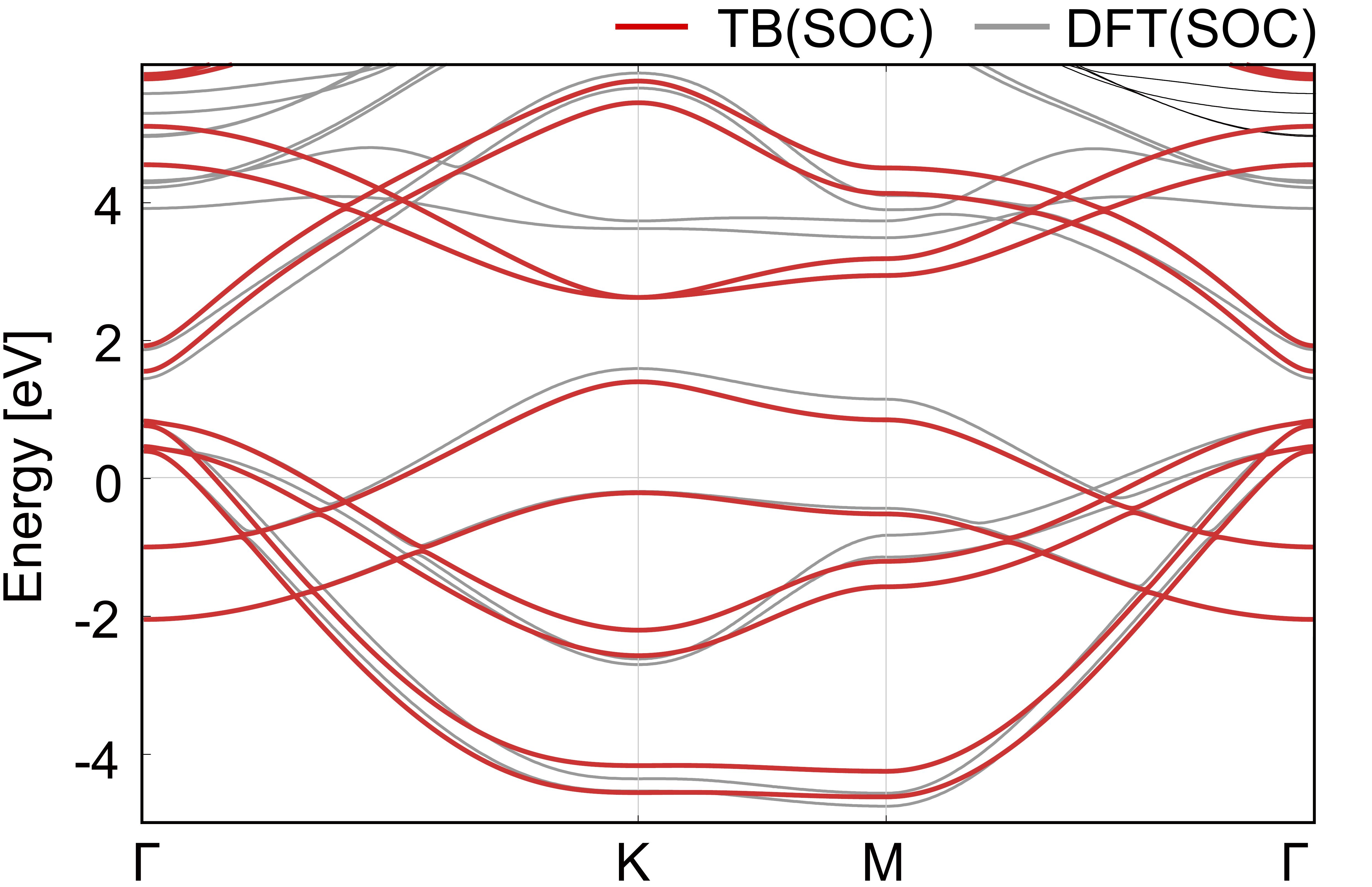}
    \caption{DFT and tight-binding (TB) electronic energy band structures with SOC. The DFT (TB) bands are colored by gray (red). The spin orientation  is set to the out-of-plane $z$-direction. 
    The following set of parameters are used for the tight-binding band structure: 
$\Delta^{s}_{\rm{C}}$ = -5.69\ eV,
$\Delta^{s}_{\rm{In}}$ = -0.99\ eV,
$\Delta^{p_{x,y}}_{\rm{C,In}}$ = 3.31\ eV,
$\Delta^{p_z}_{\rm{C}}$ = -0.62\ eV ,
$\Delta^{p_z}_{\rm{In}}$ = 2.62\ eV ,
$V_{ss\sigma}$ = -1.5\ eV,
$V_{sp\sigma}$ = 2.4\ eV ,
$V_{pp\sigma}$ = 2.8\ eV,
$V_{pp\pi}$ = -1.0\ eV,
$B^{p_z}_{\rm{C}}$ = 0.8\ eV,
$B^{p_z}_{\rm{In}}$ = 0\ eV,
$B^{p_{x,y}}_{\rm{C}}$ = 0.19\ eV,
$B^{p_{x,y}}_{\rm{In}}$ = 0\ eV,
$t^{p_z}_1$ = 1.0\ eV,
$t_{2,\rm{C}}$ = 0.1\ eV,
$t_{2,\rm{In}}$ = 0.01\ eV,
$\lambda$ = 0.034\ eV.
        }
    \label{fig:tb_bands}
  }
\end{figure}

\figr{fig:tb_bands} shows the calculated tight-binding bands compared with the DFT bands, where the effective Zeeman field is applied along the out-of-plane $z$-direction. 
We find that the next nearest-neighbor hopping between $p_z$-orbitals is indispensable for reproducing hexagonal warping of the Fermi surface. The hexagonal warping results in the dispersion of energy along the nodal lines, allowing for the formation of the electron and hole pockets.  The energy bands nicely reproduce the essential features of the first-principles results, especially the nodal structure. The symmetry analysis is also in good agreement with the first-principles results, which supports that our symmetry analysis is valid.
}

\section{Conclusion}
\red{
In summary, we have performed first-principles calculations to investigate the atomic, magnetic, and electronic structures of monolayer hexagonal indium carbide ({\em h}-InC), which realizes a stable ferromagnetic nodal-line metal. Without spin-orbit coupling, we found that doubly degenerate nodal lines are formed from spin-polarized bands. Both the type-I and type-II nodal line coexist, each of which is characterized by a distinctive pattern of the Fermi surface geometry. In the presence of SOC, an easy-plane type magnetocrystalline anisotropy arises for the ferromagnetic state. In addition, SOC provided an opportunity to tune the nodal lines to the topological Weyl points via the engineering the direction of spin-polarization, which might be potentially useful for the future device related with spins. Moreover, the unique Fermi surface topology, accompanied by the coexisting type-I and type-II nodal lines, can give rise to interesting consequence related with quantum oscillations.}

\begin{acknowledgments}
This work was supported by the Sungkyun Research Fund 2017 from Sungkyunkwan University. The computational resource was provided from the Korea Institute of Science and Technology Information (KISTI).
\end{acknowledgments}

\bibliography{refs}

\end{document}